\documentclass[12pt,preprint]{aastex}

\def\cxo{\textit{Chandra X-ray Observatory$~$}}

\begin{document}
\title{Discovery of Hot Supergiant Stars Near the Galactic Center}

\shorttitle{}

\author{Jon C. Mauerhan\altaffilmark{1}, Michael P. Muno\altaffilmark{2}, and Mark Morris\altaffilmark{1}}

\altaffiltext{1}{Department of Astronomy and Astrophysics, UCLA, 405 Hilgard Ave., Los Angeles, CA 90095}
\altaffiltext{2}{Space Radiation Laboroatory, California Institute of Technology, Pasadena, CA, 91125}

\begin{abstract}

We report new results of a campaign to find Wolf-Rayet and O (WR/O) stars and high-mass X-ray binaries (HMXBs) in the Galactic center. We searched for candidates by cross-correlating the {\textit{Two-Micron All-Sky Survey}} with a deep \cxo catalog of point sources in the Radio Arches region. Following up with $K$-band spectroscopy, we found two massive stellar counterparts to CXOGC J174555.3-285126 and CXOGC J174617.0-285131, which we classify as a broad-lined WR star of sub-type WN6b and an O Ia supergiant, respectively. Their X-ray properties are most consistent with those of known colliding-wind binaries in the Galaxy and the Large Magellanic Cloud, although a scenario involving low-rate accretion onto a compact object is also possible. The O Ia star lies 4.4 pc in projection from the Quintuplet cluster, and has a radial velocity consistent with that of the Quintuplet, suggesting that this star might have escaped from the cluster. We also present the discovery of a B2 Ia supergiant, which we identified as a candidate massive star using 8 $\micron$ $Spitzer$ maps of the Galactic center in a region near the known massive X-ray-emitting star CXOGC J174516.1-290315.  We discuss the origin of these stars in the context of evolving stellar clusters in the Galactic center. 
 
\end{abstract}


\section{INTRODUCTION}
The value of combining X-ray and infrared observations in the search for massive stars in the Galactic center has recently been demonstrated with the discovery of luminous, emission-line star counterparts ($L_{bol}\sim10^{5-6}L_{\odot}$) to X-ray sources within a projected distance of 30 pc from Sgr A*. (Muno et al. 2006b, Mikles et al. 2006). The objects found may either be massive Wolf-Rayet (WR)/O stars in colliding-wind binaries (CWBs), accreting neutron stars and black holes in high-mass X-ray binaries (HMXBs), or extraordinary isolated massive stars.  The search for objects such as these in the Galactic center is important for the following reasons.  A more complete census of the massive stellar population is crucial in understanding the mode of star formation in the Galactic center, where adverse environmental conditions such as turbulent molecular clouds, milliGauss magnetic fields and strong tidal forces may result in a relatively large Jeans mass for collapsing cloud cores (Morris 1993, Morris \& Serabyn 1996). It is unknown if these conditions result predominantly in the formation of rich clusters of massive stars such as the Arches and Quintuplet, which are among the most dense and massive clusters known in the entire Galaxy (Figer et al. 1999). Clusters such as these will be subject to the strong tidal forces of the Galactic center, and as a result will dissipate and fade from view into the dense field population of cool giants within $\sim$10 Myr (Kim, Morris \& Lee 1999; Portegies-Zwart et al. 2002, 2003). Thus, X-rays emitted by a subset of massive stars may act as a guide toward the location of such clusters. 

Of additional importance is understanding what fraction of the X-ray population currently detected within 300 pc of  Galactic center by the \cxo is composed of HMXBs.  The inner 300 pc of the Galaxy contains 1\% of the Galactic stellar mass (Launhardt, Zylka \& Mezger 2002), and so provides a statistically meaningful sample of compact objects. If star formation in the Galactic center does produce a higher fraction of massive stars and clusters than in the Galactic disk, this may result in the production of a population of Galactic center HMXBs, in excess of the number predicted to exist from population synthesis models (Pfahl, Rappaport \& Podsiadlowski 2002). These models are based upon uncertain assumptions about the angular momentum lost during the common-envelope phase, and the kicks imparted to compact objects at birth in supernovae. The detectability of HMXBs is dependent on the physics of X-ray production at the low accretion rates expected to persist in the majority of these systems, where the effect of magnetic fields in controlling the accretion flow is uncertain (Ikhsanov 2001, Liu \& Li 2006). Therefore, a comparison of numbers with the recent star formation history of the Galactic center will allow a constraint on models of HMXB formation, evolution and X-ray production. 

Recently, Muno et al (2006a) produced a new catalog of Galactic center X-ray sources. In addition to the existing shallow catalog of the inner 2$^{\circ}$$\times$0.8$^{\circ}$ (Wang et al. 2002, Muno et al. 2003), the new deeper observations of the Radio Arches region reveal an excess of bright X-ray sources over the general population of old cataclysmic variables. Although the number of  HMXBs contributing to the original X-ray population of Muno et al. (2003) has been constrained to $<$10\% (Laycock et al. 2005), it remains unknown what fraction of the current complete catalog of $>$4000 X-ray sources consists of massive WR/O stars and HMXBs. We present here new results of our IR spectroscopy campaign to identify counterparts to the X-ray population. Two massive stellar counterparts to $Chandra$ X-ray sources have been discovered. Also, an examination of archival 8 $\micron$ maps of the Galactic center from $Spitzer$ reveals interesting warm-dust structures, most likely heated and sculpted by the winds and radiation of hot stars. A careful examination of warm dust morphology can lead to the discovery of additional massive stars, as evidenced by our discovery of a B2 Ia supergiant. In this paper, we present $K$-band spectra of new massive stars, and discuss their possible origins.

\section{OBSERVATIONS}
\subsection{X-ray and Infrared-Identified Massive Stellar Candidates}
We cross-correlated the positions of infrared stars from the {\textit{Two-Micron All-sky Survey}} (2MASS) with X-ray point-source positions from $Chandra$ observations of two 17\arcmin$\times$17\arcmin$ $ fields in the Galactic center, which consisted of a $\sim$50 ks observation centered on the Radio Arches region (Muno et al 2006a), and a $\sim$100 ks observation centered on the Arches cluster (Wang et al. 2006). A search radius of 1\arcsec$ $ was used for the cross correlation, equivalent to the 1\arcsec$ $ positional error circle of $Chandra$ point sources. The X-ray data allowed the detection of Galactic center sources as faint as $L_{X}\sim1\times10^{32}$ erg s$^{-1}$, five times deeper than the shallow Galactic center survey of Wang et al. (2002). To further discriminate foreground sources, we limited our sample of candidate IR counterparts to those sources having very red IR colors ({\it H--$K_{s}$}$>$1.3), and X-ray colors indicative of an absorption column of gas and dust having $N_{H}>4\times10^{22}$ cm$^{-2}$ (see Muno et al. 2006a). We have assembled a catalog of 35 X-ray sources having bright infrared counterparts ($K_{s}<14$) that is likely to contain a number of WR/O stars and HMXBs. The {\textit{Chandra}}/2MASS cross-correlation, comparing the results of analysis on astrometrically-aligned catalogs with that of randomly offset catalogs, indicates that the number of X-ray/IR matches is in excess of that expected from random chance alone by $\sim$30\%. This is not surprising since the Radio Arches region is a known site of active star formation (Cotera et al. 1999; Figer et al. 1999). 

We also examined archival images from the mid-IR survey of the Galactic center with the {\it Spitzer Space Telescope} (Stolovy et al. 2006) in the vicinity of the known massive X-ray-emitting star CXOGC J174515.1-290315 (hereafter X174515.1 - Muno et al. 2006b).  We identified a candidate massive star near the center of curvature of a shell of 8 $\micron$ emission, shown in Figure 2, which we name SSTU J174523.1--290329, in accordance with the IAU naming convention for unregistered $Spitzer$ sources. Archival $Chandra$ observations of fields containing Sgr A*, totaling nearly 1 Msec of exposure time obtained over the course of 6 years, were used to search for an X-ray counterpart to SSTU J174523.1--290329. No X-ray counterpart was detected. Table 1 provides X-ray limits for this source.

\subsection{Infrared Spectroscopy of Candidate Massive Stars}
On the nights of July 19 and 20, 2006, we began $K$-band grism spectroscopy of our candidate sample with the Keck II telescope and NIRC2 instrument in conjunction with adaptive-optics (AO). We used the wide camera setting (0\farcs04 pix$^{-1}$) and the medium resolution grism with a 0\farcs06 slit, resulting in an effective spectral resolution of $R$$\sim$2200 ($\sim$70 km s$^{-1}$).  We observed the A1V standard star HD 19550 at the end of our observations for telluric correction. The HI Br$\gamma$ feature in the A-star spectrum was removed by interpolating the continuum over the line before telluric division. The $K$-band spectra of argon and xenon dome lamps were used for wavelength calibration. 

$K$-band spectra of a total of 6 X-ray/IR matches were obtained, and 2 of them were found to be hot emission-line stars coincident with $Chandra$ sources CXOGC J174555.3-285126 and CXOGC J174617.0-285131. The other 4 sources exhibited CO bandheads at $\lambda\approx2.3~\micron$, typical of the field population of cool giants and foreground dwarfs. Due to the extreme rarity of emission-line stars relative to the late-type field population, we assume that the emission-line stars represent the true IR counterpart to the X-ray source. In addition to the $K$-band spectra of the 2MASS counterparts to $Chandra$ sources, a $K^{\prime}$-band spectrum of SSTU J174523.1--290329 and an image of the field  were obtained. The results of spectroscopy are presented in the following section. The basic infrared and X-ray photometric properties of our candidates are given in Table 1. 
\section{ANALYSIS}
\subsection{$K$-band Spectral Comparison with Established WR and O stars}
\subsubsection{The IR counterpart to CXOGC J174555.3$-$285126}
The $K$-band spectrum of CXOGC J174555.3$-$285126 (hereafter X174555.3) exhibits very broad emission lines of HeI and HeII, indicative of a classical WR star (evolved helium-burning star) with high-velocity winds. Figure 1 (left panel) illustrates the strong similarity of the X174555.3 $K$-band spectrum to that of the central parsec cluster star IRS16SE2 (Horrobin et al. 2004, Martins et al. 2006) and the star WR134, both of which have been classified as early- nitrogen-type WR stars (WNE) of sub-type WN5-6 and WN6, respectively. In addition to the spectral comparison in Figure 1, the sub-type of X174555.3 can also be estimated as WN5-6  on the basis of the equivalent width (EW) ratios of the HeII and HeI lines, having EW$_{(2.189~\micron)}$/EW$_{(2.112~\micron)}$~$\sim$~2-4 and EW$_{(2.189~\micron)}$/EW$_{(2.166~\micron)}$~$\sim$~1-2 (see Figer, McClean \& Najarro 1997 (FMN97) -- their Figure 1). The EW measurements for X174555.3 are given in Table 2 along with the published values for the comparison star WR134. 
WN stars are designated as broad-lined, with an s or b suffix included in their sub-type, if they exhibit a 2.189~$\micron$ emission line with FWHM~$\geq$~130 \AA~(Crowther et al. 2006). The WN6 star WR134 has FWHM (2.189~$\micron$)~=~175 \AA~, reported in FMN97, and is thus given the broad designation in van der Hucht (2001). By the comparison of WR134 with X174555.3 in Figure 1, it is clear that the latter has the broadest 2.189 $\micron$ line. In conclusion, we classify X174555.3 as a WN6b star.

\subsubsection{The IR counterpart to CXOGC J174617.0$-$285131}
The $K$-band spectrum of CXOGC J174617.0$-$285131 (hereafter X174617.0) exhibits Br$\gamma$ and He I emission lines as well as the CNO product NIII.  Figure 1 (right panel) shows $K$-band spectra of X174617.0 and the comparison stars HD 152386, HD 153919 (spectra from Hanson et al. 1996), and WR 108 (spectra from Crowther \& Smith 1996). Both HD 152386 and WR 108 have been classified by Bohannan \& Crowther (1999) as WN9ha-type stars, a unique subgroup of WR star exhibiting significant hydrogen content from the Pickering-Balmer decrement (h suffix), with the upper Balmer lines showing intrinsic absorption components (a suffix).  On the other hand, HD 153919 is classified as an O Ia star (specifically O6.5 Iaf$^{+}$; Hanson et al. 1996), since the spectra of bonafide WR stars must exhibit HeII in either pure emission or P Cygni. Both HD 152386 and WR 108 show a P Cygni feature at 2.189 $\micron$, while HD 153919 and X174617.0 only show evidence of absorption at his wavelength. Furthermore, the WN9ha stars show evidence for HeI 2.058 $\micron$ P Cygni profiles, whereas X174617.0 and the O6.5 Iaf$^{+}$ star HD 153919 do not show any convincing evidence of this transition. Therefore we classify X174617.0 as an O Ia star. 

\subsubsection{SSTU J174523.1--290329}
Figure 3 shows a $K^{\prime}$ AO image of SSTU J174523.1--290329 (hereafter S174523.1). Since there was initial confusion as to which of the 2 bright stars may be responsible for the morphology of the surrounding 8 $\micron$ emission, we obtained $K^{\prime}$-band spectra of both of them. In Figure 4, the $K^{\prime}$-band spectrum of the fainter star ($K$=10.3) exhibits absorption lines of Br$\gamma$ and HeI as well as a HeI emission line at 2.058 $\micron$, bearing close resemblance to the B2 Ia supergiant HD134959 (Hanson et al. 2005). In addition to the HeI 2.058 $\micron$ emission line (only present in hot supergiants), the supergiant status of this star is further evidenced by the fact that the Br$\gamma$ and HeI 2.161 $\micron$ lines are not blended, in contrast to B main-sequence stars, where their smaller radius results in a more rapidly rotating star with rotationally broadened Br$\gamma$ lines. 

The brighter nearby star ($K$=8.7, spectrum not presented here) exhibits CO bandheads, indicative of a late-type star, perhaps a red supergiant. Due to our use of a $K^{\prime}$ filter for this particular observation, we lack the spectral coverage necessary at $\lambda$$>$2.28 $\micron$ to analyze the CO bandhead in sufficient detail to aid in determining supergiant status, although the photometry to follow in \S3.2, and the assumption of Galactic center distance, do indicate that this is a red supergiant.  

Felli et al. (2002; hereafter F02) report the location of these stars as the position of a candidate massive young stellar object (YSO), based upon photometric measurements at 7 and 15 $\micron$ with the {\it Infrared Space Observatory} ISOCAM instrument (source ISOGAL-PJ174523.1$-$290329). However, the ISOCAM beam did not resolve the brighter late-type star from the B2 Ia supergiant, and they are barely resolved in the $Spitzer$ 8 $\micron$ image (Figure 2). Figure 3 shows our $K$-band AO image of which clearly resolves these two stars. A $K$-band spectrum (not presented here) was also obtained of an additional massive YSO candidate from F02, lying 22\arcsec~ to the north-east of the B2 Ia star (source ISOGAL-PJ174523.9$-$290310) -- marked in Figures 2 and 3), exhibiting deep CO bandheads indicative of late spectral-type. 

\subsection{Extinction and Luminosity}
The extinction in the $K_{s}$-band for X174555.3, X174617.0, and S174523.1 was derived using the relation of Nishiyama et al. (2006), which gives $A_{K_{s}}=1.44E_{H-K_{s}}$ and $A_{K_{s}}=0.494E_{J-K_{s}}$ for stars located near the Galactic center. The observed $JHK_{s}$ photometry was taken from 2MASS for  X174555.3 and X174617.0 (see Table 1), and from the IRSF SIRIUS survey for S174523.1 (SIRIUS photometry provided by Shogo Nishiyama and Tetsuya Nagata). For each star, the intrinsic $(H-K_{s})_0$ and $(J-K_{s})_0$ colors of comparison stars of the same spectral type were taken from the literature and used to derive two values of $A_{K_{s}}$, which were averaged and combined with the observed photometry to produce a final value for the $K_{s}$-band extinction. 
The absolute $K$-band magnitude was then derived assuming a distance of 8 kpc to the Galactic center (DM~=~14.5). The results are summarized in Table 3. To calculate the total luminosity of each star, bolometric corrections were derived using the absolute photometry and derived luminosities of comparison stars provided in the literature (see Table 4). 

\subsection{X-ray Analysis}
The X-ray data were reduced using the methods described in detail in Muno et al. (2003). The X-ray properties of the stars of this work were determined using the ACIS EXTRACT routine from the Tools for X-ray Analysis (TARA) \footnote{www.astro.psu.edu/xray/docs/TARA/}, CIAO version 3.3 and CALDB 3.2.2. 

Neither X174555.3 nor X174627.0 were detected in the soft X-ray band (0.5--2.0 keV), indicating that these sources are highly absorbed by gas and dust, as is expected for objects located near the Galactic center. The soft color HR0=$(h-s)/(h+s)$, where $h$ and $s$ are the respective fluxes in the 2.0-3.3 keV and 0.5--2.0 keV bands, is therefore limited to HR0$>0.62$ and HR0$>0.72$ for X174555.3 and X174617.0, respectively.  The hard color HR2, where $h$ and $s$ are the respective fluxes in the 4.7--8.0 keV and 3.3--4.7  keV bands, is HR2=--0.43 (--0.74, --0.20) and HR2=--0.37 (--0.72, --0.12) for X174555.3 and X174617.0, respectively, with the 90\% confidence interval given in parentheses. The hard colors of these stars are significantly lower than the mean value of HR2=0.05 for the general source population reported in Muno et al. (2006a), and are consistent with emission from a $kT\approx2$ keV thermal plasma spectrum or a steep power-law spectrum with $\Gamma>3$ (see their Figure 3), where $\Gamma$ represents the photon index. Assuming a distance of 8 kpc to the Galactic center, the derived X-ray luminosities are given in Table 4. The B2 Ia supergiant S174523.1 was not detected in X-rays.

We also searched for X-ray variability. To search for flux variations over $\sim$month timescales,  the average background-subtracted flux from each observation was computed and compared with the null hypothesis of a constant mean flux using a $\chi^{2}$ test. The values of $\chi^{2}$ had at least a 20\% chance of being caused by random fluctuations. To search for variations on $\sim$day time scales, the arrival times of photons
were compared against the null hypothesis of a uniform distribution in time using a Kolmogorov-Smirnov
test. The values of the KS statistic had at least a 10\% chance of being caused by 
random fluctuations. We conclude that neither source varied by more than
a factor of 2.

\section{DISCUSSION}
\subsection{The nature of X174555.3 and X174617.0}
The X-ray and bolometric luminosities of X174555.3 and X174617.0 listed in Table 4 are consistent with the $L_{X}/L_{bol}=10^{-7\pm1}$ relation exhibited by both single O stars (Skinner et al. 2006a) as well as the CWBs of the Galactic starbursts Westerlund 1 (compare Crowther et al. 2006 with Skinner et al. 2006b) and NGC 3603 (Moffat et al. 2002), and the R136 cluster of the 30 Doradus region of the Large Magellanic Cloud (Portegies-Zwart, Pooley \& Lewin 2002, hereafter PPL02 -- see their Figure 4). Thus, we cannot firmly distinguish between single star and CWB scenarios using X-ray/IR photometry data alone. However, single stars are not known to be steady sources of hard X-rays and tend to have X-ray spectra dominated by a cool component with $kT\approx0.6$ keV (Oskinova 2005), while CWBs such as those in Crowther et al. (2006) and PPL02 do exhibit hard X-ray spectra with $kT$~$>$~1 keV temperatures. Therefore, we interpret the 2 keV X-ray temperature of X174555.3 and X174617.0 as a strong indication of colliding-wind binarity. 

Although far less common, it is also possible that the X-ray emission of these stars arises from  a wind accreting neutron star, which would make them quiescent HMXBs. Indeed, very few HMXBs have been observed with $L_{X}$$<$10$^{34}$ erg s$^{-1}$, most likely because of selection effects against finding them. Such low luminosities would require a low accretion rate of material, which may be prevented from reaching the neutron star's surface by a magnetic propeller. Such scenarios have been invoked to explain the quiescent, non-pulsed X-ray emission of several known transient X-ray pulsars in HMXBs, in which case the X-ray emission arises at a magnetospheric boundary in the propeller regime or from deep crustal heating from pycnonuclear reactions during outburst (Campana et al. 1998, 2002).  As stated in \S3.2, it is possible that the X-ray emission of X174555.3 and X174617.0 represents a non-thermal power-law with $\Gamma>3$, consistent with the transient HMXB 4U 0115+63 in its quiescent state, which has an observed $\Gamma=2.6_{-1.3}^{+2.1}$ (Campana et al. 2002). 

In conclusion, the observations of X174555.3 and X174617.0 are most consistent with a CWB scenario, although low accretion rate HMXB also remains a possible interpretation. Future monitoring of the IR spectra of these sources may reveal  periodic velocity variations, like those observed in the WN stars WR 134 (P=2.3 days; Morel et al. 1999), and EZ CMa (P=3.8 days; Skinner, Itoh \& Nagase 1998). Such observations could confirm the presence of companions and constrain their nature. 

\subsection{The Environment of the Blue Supergiant S174523.1}
Figure 2 shows a $Spitzer$ 8 $\micron$ image of a field containing S174523.1. The warm dust morphology in the region surrounding S174523.1 has a shell- or bubble-like appearance, indicative of a strong stellar wind. Wind-blown bubbles are predicted to surround B2 Ia stars, which have  $T_{eff}\sim20$ kK and terminal wind velocities of $v_{\infty}$=500--600 km s$^{-1}$ (Crowther, Lennon \& Walborn 2006). The physical parameters of the bubbles will be influenced by the stellar type, age, and the density and relative velocity of the ambient ISM with respect to the star (Weaver et al. 1977; Canto, Raga \& Wilkin 1996).  The bubble apparently surrounding S174523.1 is similar in appearance to some more nearby examples of stellar wind-blown bubbles from OB stars (e.g. the bubbles revealed by $Spitzer$; Churchwell et al. 2006). Assuming a distance of 8 kpc, the estimated radius of the bubble surrounding S174523.1 is $r_{bub}\sim0.7$ pc, with an outer shell thickness of $\Delta{r_{shell}}\sim0.2$ pc. This is consistent with the observations of Churchwell et al., in which bubbles of known distance have $r_{bub}$=0.1--16.9 pc and $\Delta{r_{shell}}$=0.01--3.9 pc. Furthermore, these bubbles typically have $r_{bub}$/$\Delta{r_{shell}}\sim0.25$, in good agreement with the value of 0.29 for the bubble surrounding S174523.1. 

The massive Ofpe/LBV X-ray star CXOGC J174515.1-290315 (hereafter X174515.1) reported by Muno et al. (2006b) lies 1.5$\arcmin$ to the west of the B2 Ia star. It is also surrounded by a much smaller shell of material that appears to lie on the same side of X174515.1 as the bubble surrounding S174523.1, with respect to Sgr A*. This may be indicative of shared stellar motion. As described in Weaver et al. (1977), stellar motion within the ambient ISM will produce non-spherical or parabolic bubbles. A suspicion naturally arises that X174515.1 and S174523.1 could be related, either born of the same molecular cloud or perhaps members of a tidally disrupted stellar cluster (see \S4.3). Identifying more candidate massive stars in this region and obtaining spectra of them should help shed light on this hypothesis.
  
\subsection{Cluster Drifters and Runaways}
Do the apparently isolated stars in the Galactic center have their origins in stellar clusters? Figure 5 is an 8 $\micron$ map of the inner 25\arcmin$ $ of the $Spitzer$ Galactic center survey (Stolovy et al. 2006), illustrating the spatial distribution of the currently known isolated massive stars in the region (by isolation we mean not within stellar clusters) and the positions of the stellar clusters. We refer the reader to this figure to facilitate the following discussion. Note that Figure 5 is not meant to illustrate the complete spatial distribution of isolated massive stars in the Galactic center due to the numerous selection effects involved in obtaining the current sample.

Stellar clusters in the Galactic center, such as the Arches and Quintuplet, will be disrupted by the strong tidal forces there. In the process they will lose stars and decrease in surface density until they fade into the stellar background before $\sim$10 Myr (Kim, Morris \& Lee 1999, Portegies-Zwart et al. 2003).  Within clusters, three-body interactions between binary and single stars can eject single stars with velocities of $\ga100$ km s$^{-1}$, and impart velocities of $\sim$10 km s$^{-1}$ to the remaining binary, which also may be hardened in the process (Portegies-Zwart et al. 2002, hereafter PZ02; Gualandris, Portegies-Zwart \& Eggleton 2004).  The internal velocity dispersion for the Arches and Quintuplet is roughly an order of magnitude less than their orbital velocity. Therefore stars ejected with $v\sim$~10 km s$^{-1}$ will continue to move more or less along the orbital path of the cluster, maintaining comparable velocities. In contrast, supernova-induced kicks are expected to produce runaway single stars and binaries with higher velocities (50-500 km  s$^{-1}$), and with escape trajectories in more random directions than stars that have been dynamically ejected (van den Heuvel et al. 2000, Dray et al. 2006).  These processes operating within the Arches, Quintuplet, or heretofore undiscovered clusters may play an important role in supplementing the inner Galaxy with massive stars and binaries, as well as their remnants. However, simulations of Galactic center clusters containing primordial massive binaries have yet to be performed, leaving the details and relative importance of the various ejection processes uncertain. 

The two X-ray-emitting stars of this work lie in close proximity to the Arches and Quintuplet clusters. X174555.3 (WN6b) lies 5.1 pc in projection from the Arches and 10.9  pc in projection from the Quintuplet, while X174617.0 (O Ia) lies 14.4 pc in projection from the Arches and 4.4 pc in projection from the Quintuplet.  The proximity raises the question of whether or not X174555.3 and X174617.0 are escapees from one of the two clusters. We suspect that X174555.3 and X174617.0  are more likely to be former members of the Quintuplet cluster, rather than the Arches, for several reasons: 1) the Quintuplet is older than the Arches (4$\pm$1 vs. 2.5$\pm$0.5 Myr, respectively -- Figer, McClean \& Morris 1998 (FMM98), Figer et al. 2002 (F02)), having had more time to evolve dynamically; 2) WNE stars such as X174555.3 have ages 3-5 Myr (Meynet \& Maeder 2005), which is consistent with the Quintuplet but too old for the Arches; 3) The O Ia star X174617.0 is only 4.4 pc in projection from the Quintuplet and has a spectral doppler shift indicating a velocity of +180$\pm$70 km s$^{-1}$, much more consistent with the +130 km s$^{-1}$ radial velocity of the Quintuplet (FMM98) than with the +95 km s$^{-1}$ radial velocity of the Arches (F02). Therefore, we suggest that X174555.3 and X174617.0 may have originated in the Quintuplet cluster, although an isolated formation scenario cannot be ruled out. Quite possibly, the star formation event which produced the Quintuplet could have been accompanied by a `aureola'\ of isolated star formation (perhaps induced star formation), leading to the production of coeval stars unbound to the cluster.  

Could the other isolated massive stars in the Galactic center also be former members of the Arches or Quintuplet clusters? Adopting 10 km s$^{-1}$ as a typical cluster ejection velocity implies that escaped stars could have drifted 25 pc and 40 pc from the Arches and Quintuplet, respectively. These distances encompass many of the known isolated emission-line stars of the Galactic center, which raises the question of cluster origin.  The  H1-H8 HII regions (Yusef-Zadeh \& Morris 1987; Zhao et al. 1993; Cotera et al 1999) could be produced by former Arches or Quintuplet members simply passing through a particularly dense region of the Galactic center ISM.  In addition to their spatial proximity, the WC8-9 and WN10 stars reported in Homeier et al. (2003), as well as the WN6-7 star SgrA-A star reported in Cotera et al. (1999), could be regarded as Quintuplet relatives on the basis of evolutionary similarities to the Quintuplet WC and WN stars (FMM98), since WR stars only spend $\approx$~few~$\times$~10$^{5}$~yr in the WNE and WC phases (Meynet \& Maeder 2005). In contrast, the Ofpe/WN9 star G0.10+0.20 of Cotera et al. (1999) could be an escapee from the Arches cluster, as evidenced by its close proximity to the Arches ($\sim$1.7 pc in projection), and by its near-infrared spectrum which indicates that it has the same evolutionary status as several of the Arches members (F02). 
 
\section{PERSPECTIVES}
Our success in employing the deep X-ray sample of the radio Arches region of Muno et al. (2006a) (17\arcmin$\times$17\arcmin$ $) in our search for WR/O stars and HMXBs is a prelude to a more extended effort to find similar objects within the inner 2$^{\circ}$$\times$0.8$^{\circ}$ of the Galactic center. This area is currently being covered in our $Chandra$ Legacy survey for deep observations of the inner 100 pc of the Galaxy, and we plan to continue our IR spectroscopy campaign to search for counterparts. As more X-ray-emitting massive stars are discovered, IR spectral monitoring could help confirm the presence of companions in suspected CWBs and HMXBs, and constrain their mass functions. Furthermore, long-term comparative proper motion studies with adaptive-optics (AO), coupled with radial velocity measurements, could reveal stellar cluster escapees through their space velocity distribution, which may also give information on the energetics of the events that ejected them, and on the tidal field of the Galactic center.

\acknowledgements
We are grateful to the anonymous referee for very insightful and helpful comments. We also thank Margaret Hanson, Paul Cowther, Fabrice Martins and Frank Eisenhauer for kindly supplying us with $K$-band spectra of hot stars for our comparative analysis, and Tetsuya Nagata and Shogo Nishiyama for sharing near infrared photometry data from the IRSF SIRIUS survey. J.~C.~M. would like to thank Simon Portegies-Zwart for a helpful interchange regarding the dynamical evolution of compact clusters in the Galactic center, and Andrea Stolte for an informative discussion regarding the kinematics of the Arches cluster. This research was supported by an NSF grant (AST-0406816).

\clearpage
\begin{deluxetable}{lcccccccc}
\tablecolumns{9}
\tablewidth{0pc}
\tabletypesize{\scriptsize}
\tablecaption{Infrared and X-ray Observational Properties of New Hot Stars\label{tab:prop_01}}
\tablehead{
\colhead{Name~~~~~~} & \colhead{R.A. (J2000)} & \colhead{Dec. (J2000)} & \colhead{$J$} & \colhead{$H$} & \colhead{$K_s$} & \colhead{$F_X$\tablenotemark{b}} & HR0\tablenotemark{c} & HR2\tablenotemark{c} \\
\colhead{} & \colhead{(deg)}  & \colhead{(deg)} & \colhead{} & \colhead{} & \colhead{} & \colhead{} & \colhead{} & \colhead{} 
}
\startdata
CXOGC J174555.3-285126 & 266.48068 & --28.85734 & 14.76 & 12.40$\pm$0.06 & 10.84$\pm$0.06  & 936.1 & $>0.62$ & --0.43 \\ 
CXOGC J174617.0-285131 & 266.57123 & --28.85876 & 14.85$\pm$0.05  & 11.77$\pm$0.02 & 10.13$\pm$0.02 &  650.3 & $>$0.72 & --0.37 \\ 
SSTU J174523.1-290329\tablenotemark{a}  & 266.34630 & --28.05814 & 16.7$\pm$0.03 & 12.7$\pm$0.02 & 10.3$\pm$0.02 & $<$6.8 & \nodata & \nodata 
\enddata
\tablecomments{Positions are accurate to 0\farcs3.  Uncertainties on the 
infrared magnitudes are provided where they are available from the 2MASS 
catalog.}
\tablenotetext{a}{IR photometry from IRSF SIRIUS survey kindly provided by Tetsuya Nagata and Shogo Nishyama.}
\tablenotetext{b}{X-ray flux given in units of 10$^{-6}$ counts s$^{-1}$}
\tablenotetext{c}{As described in the text, X-ray color is defined as $(h-s)/(h+s)$, where soft color (HR0) is represented when $h$ and $s$ are the respective fluxes in the 2.0--3.3 keV and 0.5--2.0 keV energy bands, and hard color (HR2) when $h$ and $s$ are the respective fluxes in the 4.7--8.0 keV and 3.3--4.7 keV energy bands.}
\end{deluxetable}

\begin{center}
\begin{deluxetable}{lcccccccc}
\tablecolumns{9}
\tablewidth{\linewidth}
\tabletypesize{\scriptsize}
\tablecaption{Equivalent Widths of New Hot Stars and Known Analogs \label{tab:prop_02}}
\tablehead{
\colhead{Star} & \colhead{~Type}  & \colhead{He I} & \colhead{C III/N III} & \colhead{He I/NIII} & \colhead{He I//H I} & \colhead{He II} & \colhead{N III} & \colhead{He II} \\
\colhead{} & \colhead{} & \colhead{$2P-2S$} & \colhead{8-7} & \colhead{$4S-3P$ } & \colhead{7-4, 7-4} & \colhead{10-7} & \colhead{} & \colhead{13-8} \\
\colhead{ } & \colhead{ }  & \colhead{2.058} & \colhead{2.104} & \colhead{2.112} & \colhead{2.161, 2.166} & \colhead{2.189} & \colhead{2.247} & \colhead{2.347} }
\startdata
\bf{X174555.3} & WN6b  & \nodata & \nodata & 49 & \multicolumn{2}{c}{185 (2.16-2.19 $\micron$ blended) } & \nodata & 37 \\
WR 134\tablenotemark{a} & WN6b  & \nodata & \nodata & 41 & 91 & 135  & \nodata & 63  \\
\bf{X174617.0\tablenotemark{d}} & O Ia  & \nodata & 1 & 5.8 & 13 & 1.2 abs & 1.9 & \nodata  \\
HD 152386\tablenotemark{b, d} & WN9ha & 4.1$\pm$0.5 & 1 $$  & 7.7$\pm$0.5 & 17$\pm$1 & em+abs & \nodata & \nodata \\
HD 153919\tablenotemark{b, c, d} & O6.5 Iaf$^+$  & 3.0 abs & $<$1 & \nodata & 7.6  & 0.9 abs & \nodata & \nodata\\
WR 108\tablenotemark{a} & WN9ha & 6 & 4 & 17 & 25 & $<$3 & 6 & $<2$ \\
\bf{S174523.1\tablenotemark{e}} & B2 Ia  & 1.6 & \nodata & 1.9 abs  & 1.33 abs, 4.6 abs & \nodata & \nodata & \nodata \\
HD 134959\tablenotemark{e, f, g} & B2 Ia  & 1.0 & \nodata & 2.3 abs & 0.9 abs, 4.5 abs & \nodata & \nodata & \nodata\\
\enddata
\tablecomments{The new stars of this work are in bold face text. Equivalent widths are given in \AA$ $. Uncertainty for the infrared line measurements is $\sim$ 0.5 \AA$ $ for X174555.3, X174617.0, and S174523.1, estimated from measurements of continuum noise in our spectra. Absorption lines are indicated by abs. All other lines are in emission.}
\tablenotetext{a}{Values from Figer, McClean, and Najarro (1997)}
\tablenotetext{b}{Values from Hanson et al. (1996)}
\tablenotetext{c}{HD 153919 is a member of the high-mass X-ray binary system V$^{*}$~V884 Sco.}
\tablenotetext{d}{Also contains HeII (2.1185 $\micron$) absorption line with EW$\sim$0.9. HD 152386 also shows emission in this region. These stars also contain weak CIV emssion at (2.069 $\micron$), although it is strongest in HD 153919 with EW=3.5$\pm$1}
\tablenotetext{e}{Also has weak Mg II emission (EW$\sim$1) at 2.137 $\micron$, and at 2.143 $\micron$ for HD 134959. There is also a weak He I absorption line EW$<$0.5 at 2.150 $\micron$.}
\tablenotetext{f}{Values from Hanson et al. 2005}
\tablenotetext{g}{Also contain weak HeI absorption at 2.182, 2,184 $\micron$}
\end{deluxetable}
\end{center}

\begin{deluxetable}{lcccccccccc}
\tabletypesize{\scriptsize}
\tablecolumns{9}
\tablewidth{0pc}
\tablecaption{Derivation of Absolute Photometry for New Hot Stars}
\tablehead{\colhead{Star} & \colhead{$(J-K_{s})_{0}$\tablenotemark{a}} & \colhead{$(H-K_{s})_{0}$\tablenotemark{a}} & \colhead{$J-K_{s}$} & \colhead{$H-K_{s}$} & \colhead{$A^{J-K_{S}}_{K_{S}}$}& \colhead{$A^{H-K_{S}}_{K_{S}}$} & \colhead{$\overline{A_{K_{S}}}$} & \colhead{$M_{K_{s}}$}   
}
\startdata
X174555.3 & 0.36 & 0.26 & 3.92 & 1.56 & 1.76 & 1.87 & 1.82 & --5.48 \\
X174617.0 & --0.11 & --0.10 & 4.72 & 1.64 & 2.39 & 2.51 & 2.45 & --6.82  \\
S174523.1 & --0.03 & 0.00 & 6.4 & 2.4 & 3.18 & 3.46 & 3.32 & --7.52
\enddata
\tablenotetext{a}{The WN6b star WR134 was used to represent the intrinsic colors of X174555.3 (Crowther et al. 2006). Synthetic photometry of an O6.5 Ia star were used for the intrinsic colors of X174617.0 (Martins \& Plez 2006). Intrinsic colors for the B2 Ia star were taken from Winkler 1997.}
\end{deluxetable}

\begin{deluxetable}{lcccccc}
\tabletypesize{\scriptsize}
\tablecolumns{7}
\tablewidth{0pc}
\tablecaption{Bolometric and X-ray Luminosity for New Hot Stars\label{tab:lxlbol}}
\tablehead{
\colhead{Star} & \colhead{Spectral} & \colhead{$BC_{K}$\tablenotemark{a}}& \colhead{$M_{bol}$} & \colhead{$L_{\rm Bol}$} &
\colhead{$L_{\rm X}$\tablenotemark{b}} &  \colhead{$L_{\rm X}/L_{\rm Bol}$}  \\
\colhead{} & \colhead{Type} & \colhead{} & \colhead{} & \multicolumn{2}{c}{(log$[$ erg s$^{-1}$$]$)} & 
\colhead{(log)} 
}
\startdata
{\bf X174555.3} & WN6b & --4.07 & --9.55 & 39.31 & 32.6 & --6.5  \\
{\bf X174617.0} & O Ia & --3.29 & --10.11 & 39.53 & 32.6 & --6.9 \\
{\bf S174523.1} & B2 Ia & --1.88 & --9.40 & 39.26 & $<$30.9 & $<$--8.7 
\enddata
\tablenotetext{a}{The $K$-band bolometric corrections and magnitudes were derived combining the derived $M_{K_{s}}$ values of Table 3 with the stellar luminosities of the following comparison stars provided in the literature: WR134 (WN6b -- Hamman, Gr\"afener \& Liermann 2006), a synthetic O6 Ia star (Martins, Schaerer \& Hillier 2005), and HD 134959 (B2 Ia --  Crowther et al. 2006, Winkler 1997).}
\tablenotetext{b}{The X-ray luminosities for these sources are 
extrapolated from the observed $Chandra$ values into the 0.5--8.0 keV band. The 0.5--8.0 keV luminosities of individual massive stars are usually dominated by $kT \approx 0.5$ keV plasma that would be undetectable through the absorption toward the Galactic center; at most 10\% of $L_{\rm X}$ should be produced by hotter $kT>$1 keV plasma. Therefore, if the X-rays originate in the winds of individual stars, then the X-ray luminosity values could uncertain by roughly an order of magnitude. On the other hand, if a substantial fraction of the X-rays are generated in the wind collision zone of a binary, then the spectra will be dominated by a harder X-ray component ($kT>1$~keV) resulting in a more accurate extrapolated value of $L_{X}$.}
\end{deluxetable}

\clearpage

\begin{figure}
\epsscale{0.75}
\plotone{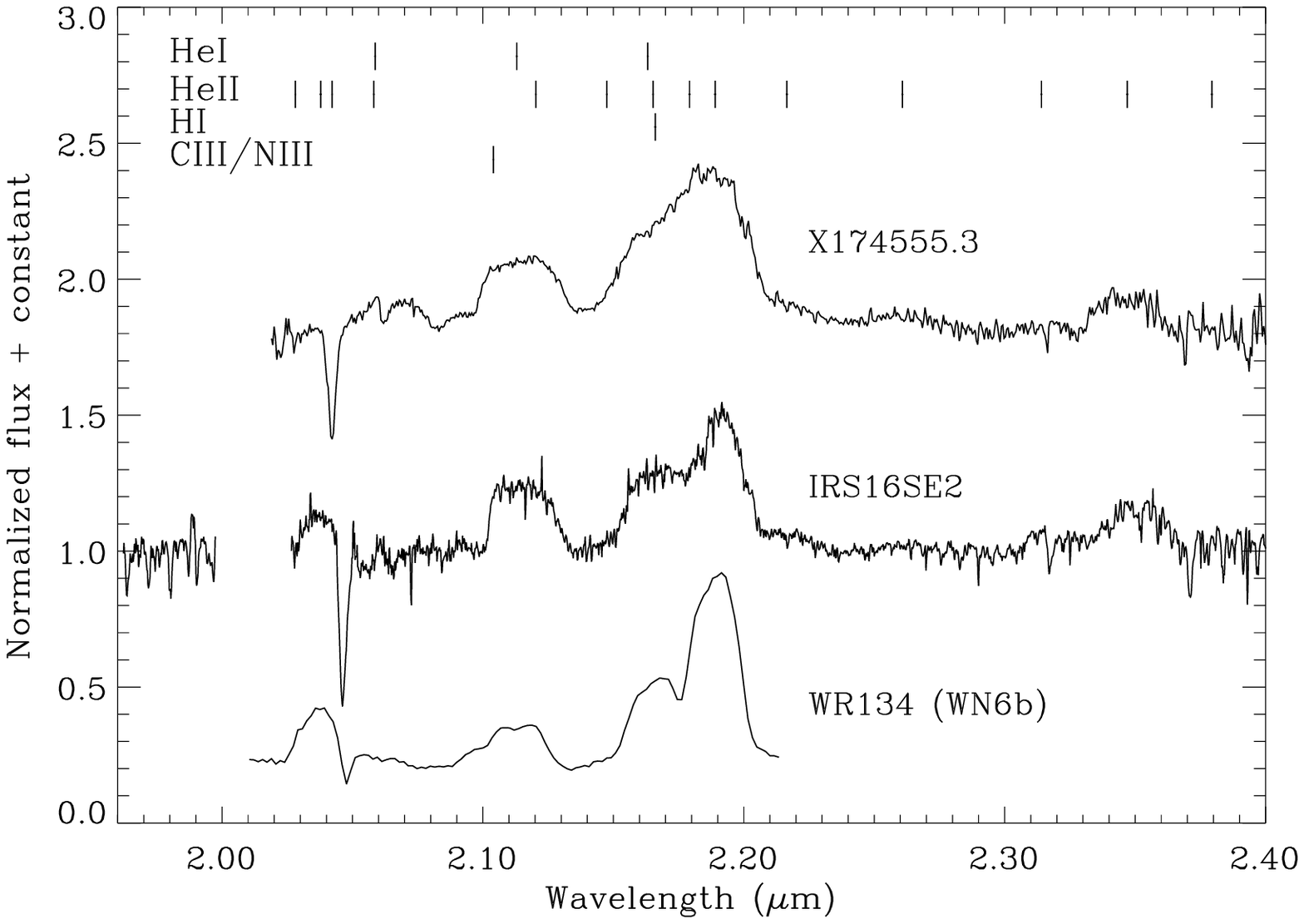}
\plotone{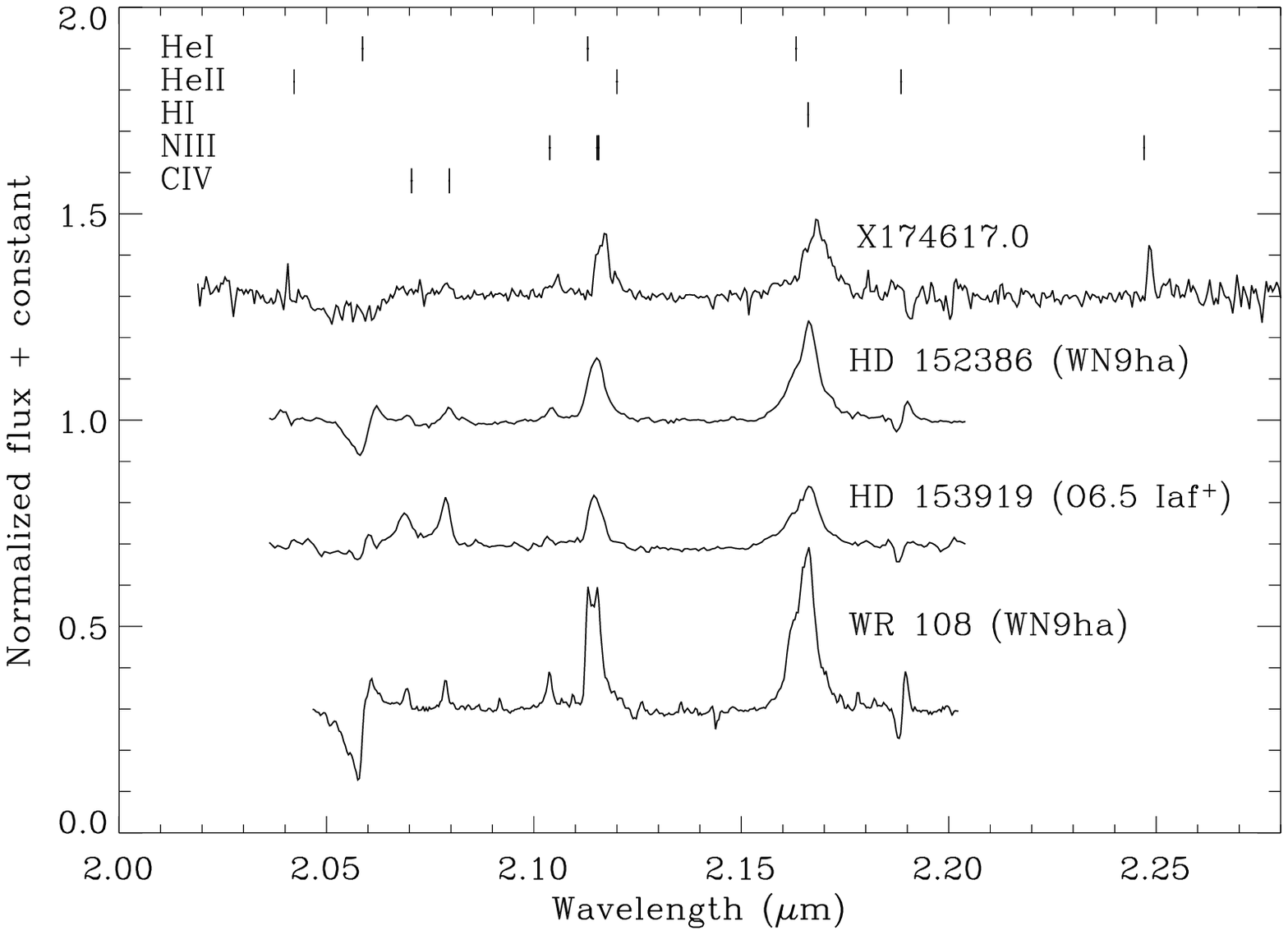}
\caption{$K$-band spectra. {\textit{Left:}} the counterpart to CXOGC J174555.3$-$285126 plotted with the WN5-6 comparison star IRS16SE2 (Martins et al. (2006) -- spectrum provided by the author), and WR134 (spectrum provided by Paul Crowther) \textit{Right:} CXOGC J174617.0$-$285131 plotted with comparison stars HD 152386 (WN9ha) and HD 153919 (O6.5 Iaf$^{+}$) from Hanson et al. (1996), and the WN9ha star WR 108 from Bohannan \& Crowther (1999).}
\end{figure}

\begin{figure}
\plotone{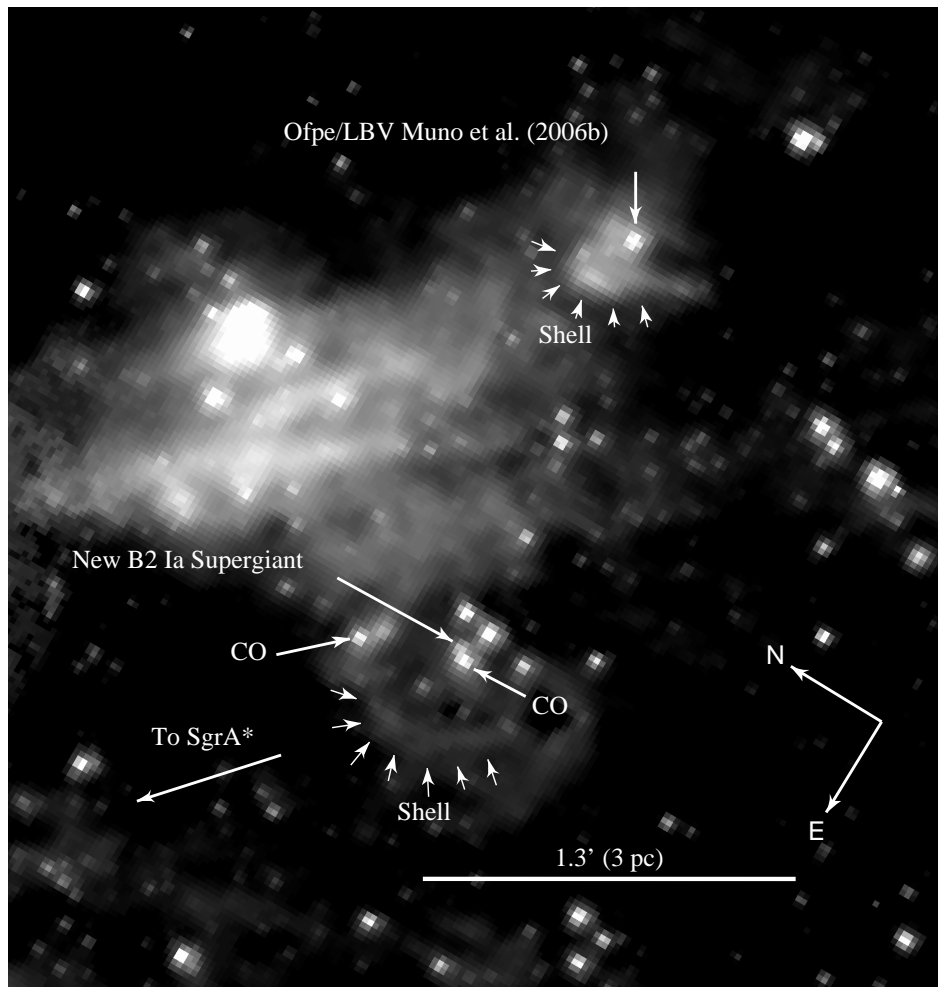}
\caption{{\textit{Spitzer}} 8$\micron$ map of the region surrounding the B2 Ia supergiant SSTU J174523.1$-$290329 (below center) and the Ofpe/LBV star from Muno et al. (2006b). Sources that we have identified as late-type stars are marked CO (see \S3.1.3). The morphology of the warm dust surrounding the B2 Ia star is indicative of strong stellar wind activity. The short arrows trace the dust shells discussed in \S4.2. The direction to SgrA* is 4.6$\arcmin$ in the lower left direction, indicated by the arrow. The image is orientated with respect to the Galactic plane. See Figure 5 to view this region in the broader context of the Galactic center.} 
\end{figure}

\begin{figure}
\plotone{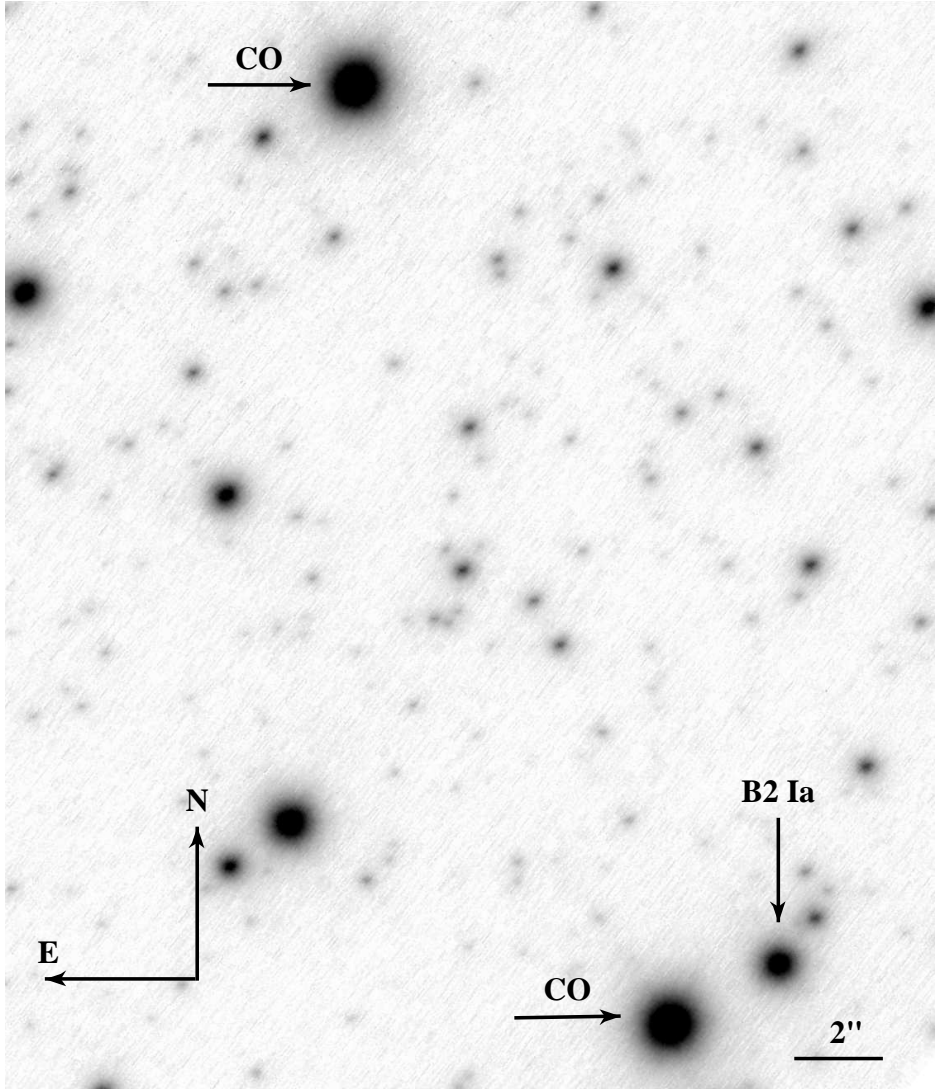}
\caption{Portion of a Keck NIRCII+AO $K^{\prime}$ image of the field containing the B2 Ia supergiant SSTU J174523.1.--290329. The sources marked CO are late-type stars exhibiting $K$-band spectra with CO bandheads (spectra not presented in this paper). }
\end{figure}

\begin{figure}
\centerline{\includegraphics[width=\linewidth]{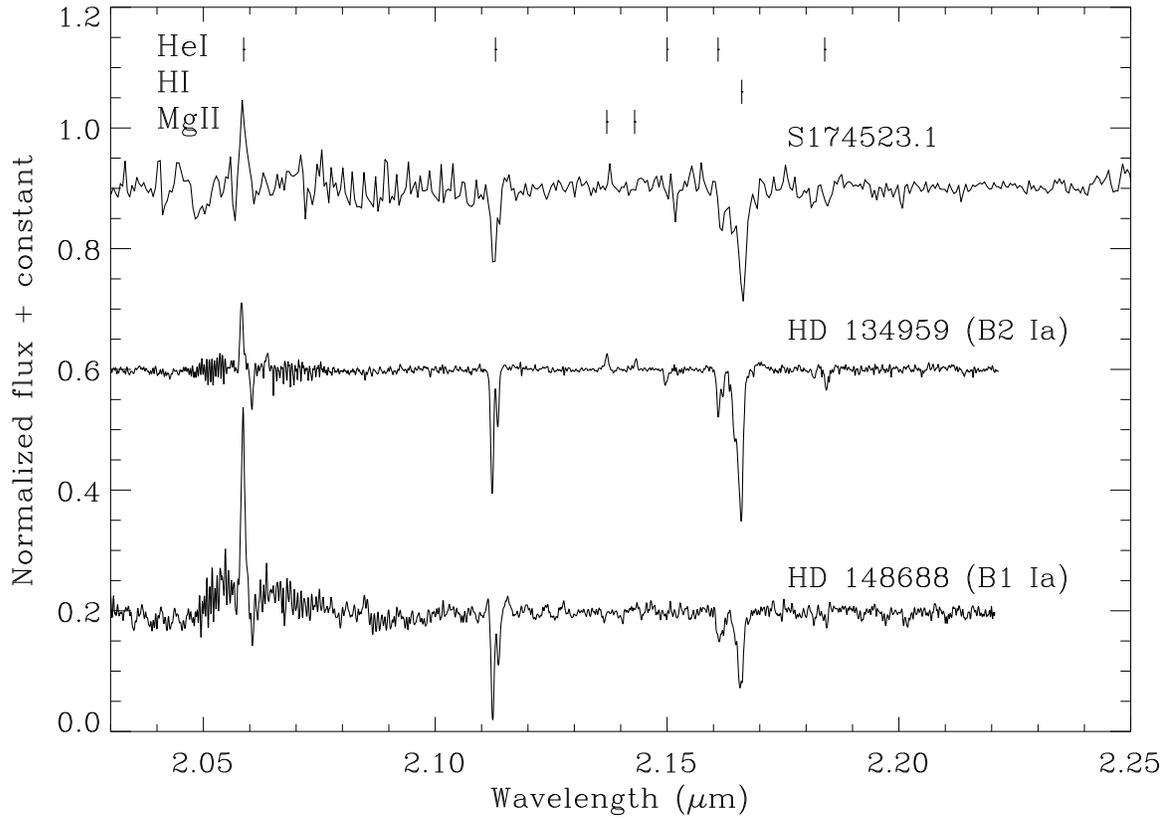}}
\caption{$K$-band spectra of the counterpart to SSTU J174523.1--290329 (S174523.1) plotted with two comparison B Ia stars from Hanson et al. 2005. The comparison suggests B2 Ia spectral type for S174523.1.}
\end{figure}

\begin{figure}
\centerline{\includegraphics[width=\linewidth]{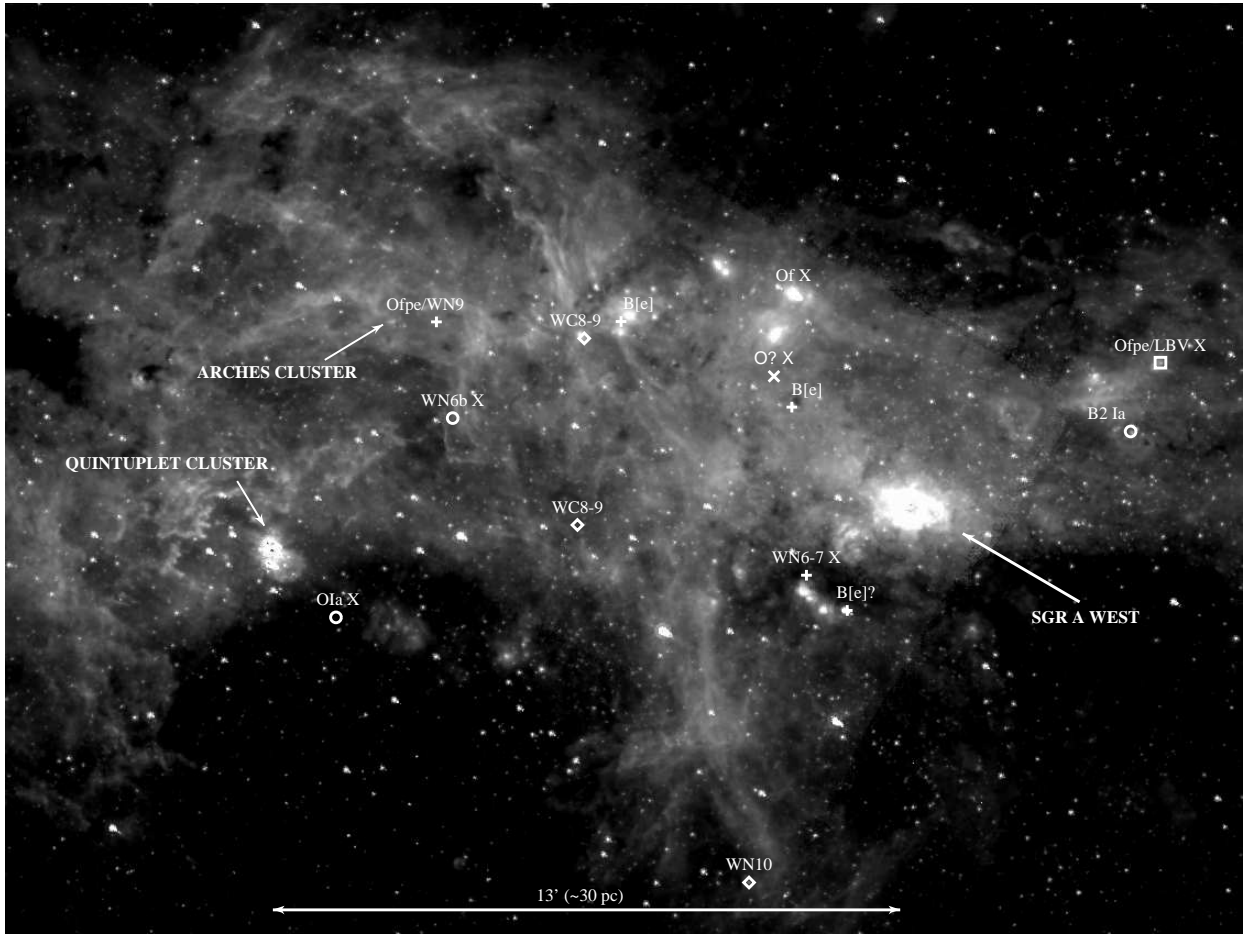}}
\caption{\textit{Spitzer} 8 $\micron$ map (logarithmic contrast stretch) of the inner 25\arcmin$ $ of the Galaxy (Stolovy et al. 2006). The positions of the stellar clusters and all known isolated emission line stars are shown with stellar classifications: Cotera et al. 1999 ({plus signs}), Homeier et al. 2003 ({diamonds}), this work ({circles}), Muno et al. 2006b ({box}), Mikles et al. 2006 ({cross}). Stars detected in X-rays are followed by an X. The image is orientated with respect to the Galactic plane.}
\end{figure}

\end{document}